\documentclass[aps,prl,twocolumn,superscriptaddress]{revtex4-1}
\usepackage[utf8]{inputenc}
\usepackage{graphicx}
\usepackage{dcolumn}
\usepackage{bm}
\usepackage{hyperref}
\usepackage{mathptmx}
\usepackage{float}
\usepackage{adjustbox}
\usepackage[T1]{fontenc}
\usepackage{xcolor}
\usepackage{amsmath}
\usepackage{amssymb}
\DeclareGraphicsExtensions{.pdf,.png,.jpg,.gif}
\graphicspath{{figures/}}

\begin{document}

\preprint{APS/123-QED}

\title{Terahertz annular antenna driven with a short intense laser pulse}

\author{N. Bukharskii}
\affiliation{National Research Nuclear University MEPhI, 31 Kashirskoe shosse, 115409 Moscow, Russian Federation}

\author{Iu.~Kochetkov}
\affiliation{National Research Nuclear University MEPhI, 31 Kashirskoe shosse, 115409 Moscow, Russian Federation}

\author{Ph.~Korneev}
\email{ph.korneev@gmail.com}
\altaffiliation[Also at ]{P.N. Lebedev Physical Institute of RAS, 53 Leninskii Prospekt, 119991 Moscow, Russian Federation}
\affiliation{National Research Nuclear University MEPhI, 31 Kashirskoe shosse, 115409 Moscow, Russian Federation}

\begin{abstract}
Generation of terahertz radiation by an oscillating discharge, excited with short laser pulses, may be controlled by geometry of the irradiated target. In this work, an annular target with a thin slit is considered as an efficient emitter of secondary radiation when driven by a short intense laser pulse. Under the irradiation, a slit works as a diode, which is quickly filled by dense plasma, closing the circuit for a travelling discharge pulse. Such a diode defines the discharge pulse propagation direction in a closed contour, enabling its multiple passes along the coil. The obtained oscillating charge efficiently generates terahertz waves with a maximum along the coil axis and controllable characteristics.     
\end{abstract}

\maketitle



Generation of electromagnetic radiation in the terahertz range (THz radiation) is rapidly advanced over the last three decades \cite{Dhillon2017}. A great interest is motivated by a diverse and rich field of foreseen applications. Due to a low photon energy the THz radiation is non-ionizing, but is strongly absorbed by inter-molecular bonds and can be used for medical imaging \cite{Qiushuo2017}, high signal-to-noise ratio pulsed spectroscopy e.g. for detection of specific chemicals \cite{Giles2008, Liu2018}. THz radiation can be used for a narrow angle secure high-speed data transmission \cite{Jianjun2018}, reaching up to several terabits-per-second, surpassing recent 5g technology \cite{Huang2011}. Intense THz are used for ultra-fast electromagnetic switching \cite{Balos2020, Kampfrath2011} and electron acceleration, characterization \cite{Downer2018} and manipulation \cite{Dongfang2018, Dongfang2020}, study of the nonlinear properties of materials \cite{Liu2012} and many other applications.


Achievable intensity is a key parameter of THz sources. With large aperture photo-conductive antennas (LAPCA) asymmetric quasi half-cycle THz pulses are generated with high stability and conversion efficiency up to several percent \cite{Yang2014}, but the field peak value is only few hundreds of kV/cm at low THz frequencies \cite{Ropagnol2016}. With birefringent crystals and the optical rectification method, single to multi-cycle $0.1...6$ THz pulses are obtained with the peak up to several MV/cm \cite{Ovchinnikov2018, Sitnikov2020} and the efficiency up to several percent \cite{Huang2013, Hauri2011}. 
Generally, the non-plasma based techniques have a limitation on the maximum intensity because of the medium breakdown. Laser plasma THz sources may be more powerful, even at very high laser intensities of $10^{17}-10^{20}$~W/cm$^2$ the THz yield does not experience any saturation \cite{Gopal2013}. It is expected that THz field strength can reach $\sim1$ GV/cm \cite{Dechard2018} and even more. 
Generation of THz radiation in laser-plasma interaction is an emerging research field. It accompanies the laser-excited wakefield \cite{Liao2015}, Cherenkov wake radiation in magnetized plasmas \cite{Bakhtiari2017}, the transient electrons currents excited either in the target bulk \cite{Li2016, Liao2016} or outside the target \cite{Herzer2018}, the coherent transition radiation (CTR) and the TNSA effects \cite{Wu2013}. Recently, transient phenomena in laser-driven wires were discussed in this context \cite{Teramoto2018, Zhuo2017, Nakajima2017,  Tian2017, Zeng2020}.  

This work presents a concept of an efficient, controllable and high-power THz radiation source based on a discharge electromagnetic pulse travelling in a solid shaped target. Within certain geometrical constraints, such targets form an emitting antenna with a relatively high Q factor due to formation of a closed circuit just after laser irradiation. Numerical modelling shows that a controllable directed narrow-band THz radiation pulse may be obtained with a high conversion efficiency. With adjustment of the interaction parameters in the presented simulations, the expected efficiency reaches several percents with the peak electric field of {$\sim 0.1$~GV/cm} at a distance of $1$~cm from the target centre, when a driver is $24$ fs laser pulse with the focal spot intensity $10^{21}$~W/cm$^2$.
\begin{figure}
    \centering
    \includegraphics[scale = 0.25]{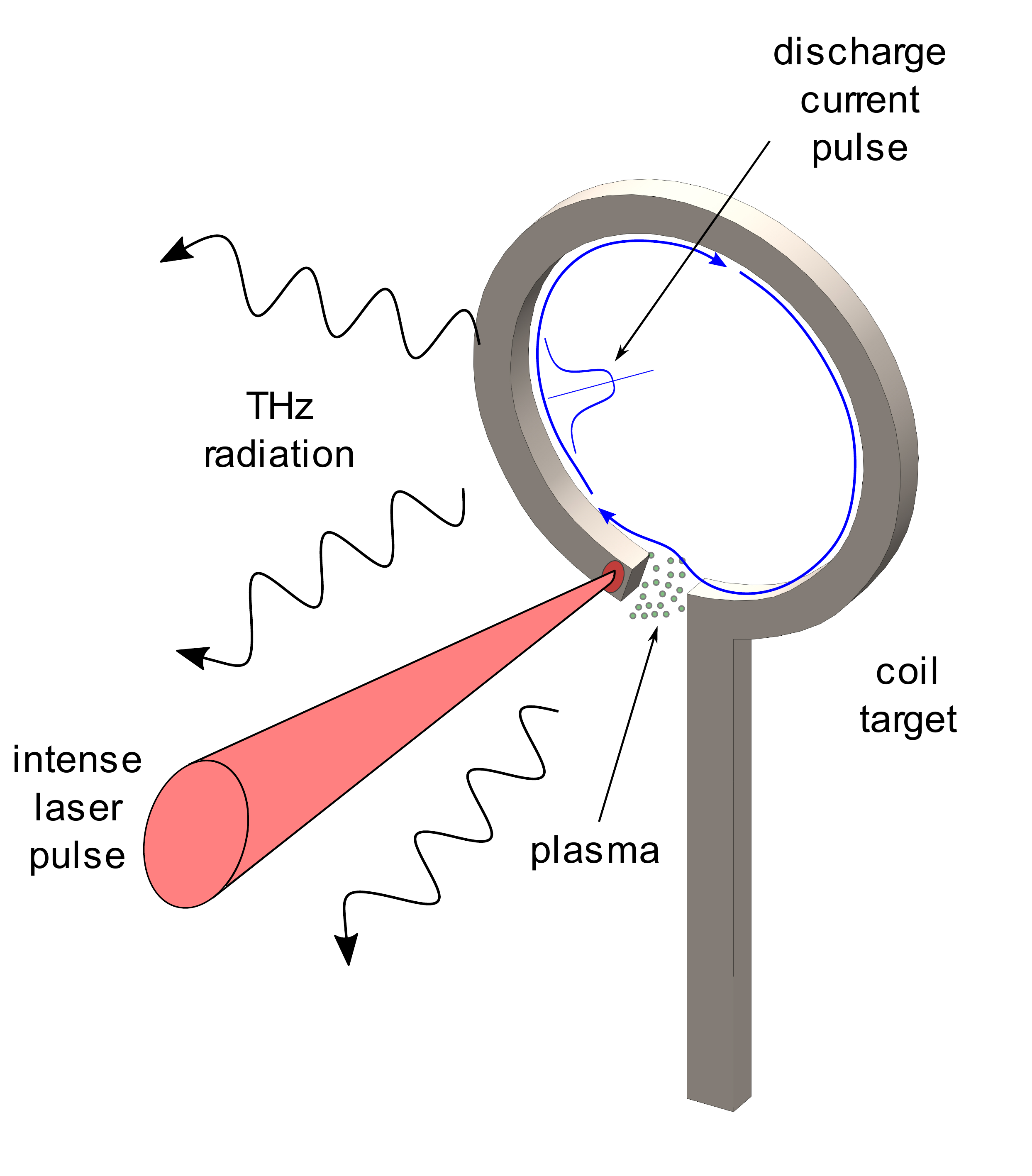}
    \caption{A sketch of the setup for generation of high-power THz radiation by laser irradiation of coil targets.}
    \label{THz_generation_setup}
\end{figure}

The principal setup is presented in Figure~\ref{THz_generation_setup}. An intense short laser pulse is focused on the open end of a miniature coil target, causing plasma formation and hot electron generation. The escaped negative charge form a strong positive potential in the interaction region. As this stage is very fast, the gap between the free coil end and the opposite side of the coil stays open, and an excited discharge current pulse propagates along the coil in the direction from the end, clockwise in Fig.~\ref{THz_generation_setup}. 
For a short driver pulse $t_p \ll {P}/{c}$, where $P$ is the coil perimeter and $c$ is the light velocity, the discharge pulse is well localized on the coil scale. While this pulse propagates along the coil, the hot plasma expands from the irradiated region and may fill the gap, electrically closing the coil. In case of the gap, closed with a dense plasma, when the discharge pulse passes the whole turn, it divides into two parts. One continues to propagate along the target stalk and the other one propagates along the coil through the plasma in the gap. Then, a fraction of the discharge current pulse makes another turn and the process is repeated for a certain number of cycles, until the trapped pulse degrades. A sufficiently narrow discharge pulse, travelling along the coil, radiates in a spectral range around the main frequency, defined as the inverse time of a single current pulse travel along the coil. The pulse propagation velocity is normally close to the light velocity, though somewhat slower, so the main radiation frequency is $\omega_0\sim 2\pi c/P = c/a$, where $a$ is the radius in the case of a circular coil.  
\begin{figure}
    \centering
    \includegraphics[width = \linewidth]{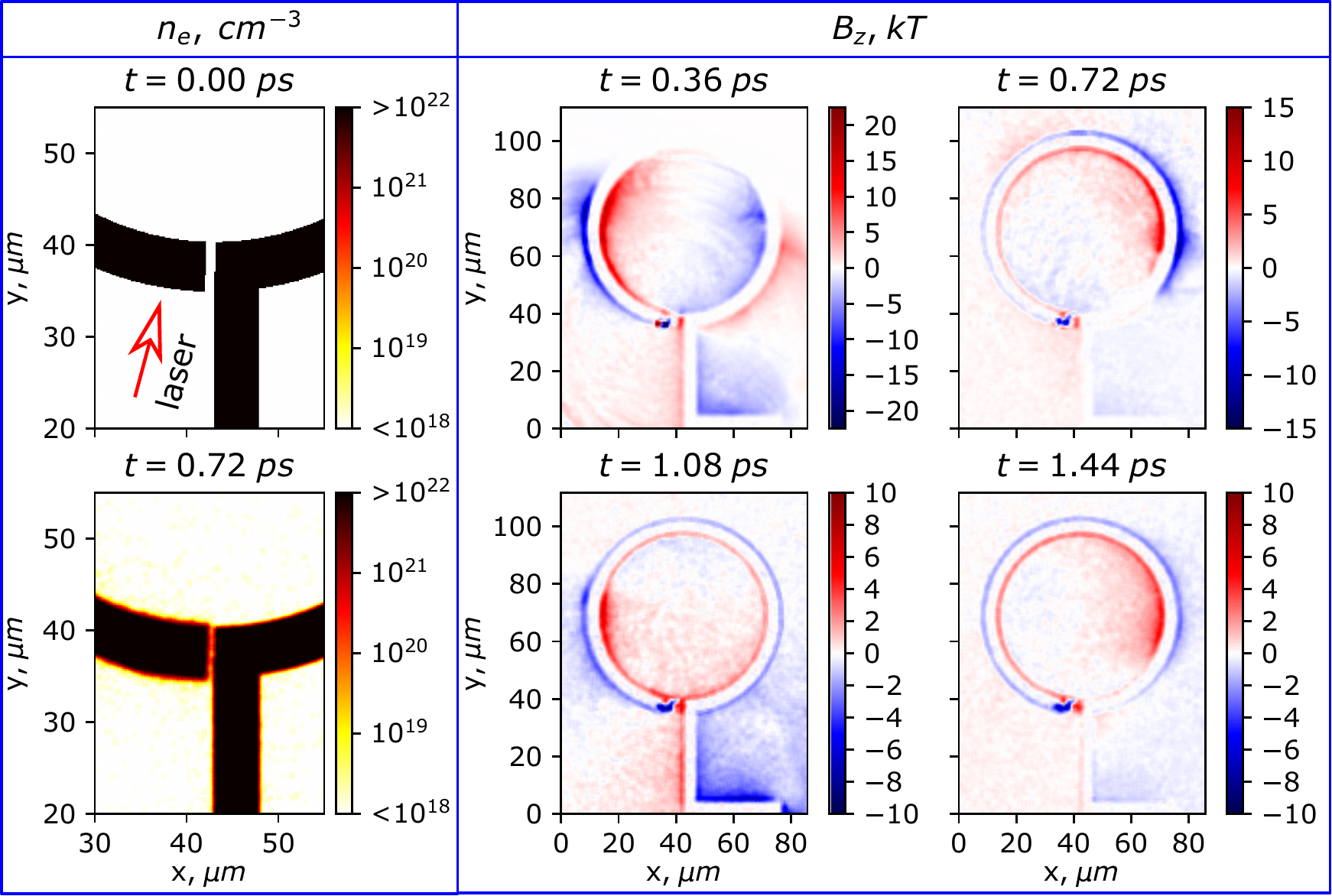}
    \caption{Results of PIC simulations for 68 $\mu$m wide 'copper' target with 0.8 $\mu$m gap.}
    \label{PIC_results}
\end{figure}

The considered setup was studied numerically using 2D Particle-in-Cell (PIC) simulations with an open source collaborative code Smilei~\cite{Smilei}. Several targets, different in geometries and materials, were studied. Target (\textbf{a}) has an outer diameter of $68~\mu$m, coil thickness of $5.6~\mu$m and gap between the coil end and its opposite side of $0.8~\mu$m, target (\textbf{b}) is the same as (\textbf{a}), but with the gap of $4~\mu$m, target (\textbf{c}) has an outer diameter of  $41~\mu$m and a gap of $0.8~\mu$m. 

The plasma that was initialized at the start of the simulation consisted of electrons and ions with electric charge $Z=10e$ and atomic mass $A=4896$ corresponding to the atomic mass of copper multiplied by a factor $76.5$ to partially compensate the decreased values of ion and electron densities in comparison to the solid state values. Additionally, 'aluminium' ($A=27 \cdot 76.5$) and 'gold' ($A=197 \cdot 76.5$) targets were studied to test how different ion masses affect the expansion of plasma near the irradiation point and consequently the possibility of the discharge current pulse propagation through the gap. The electron density was $\approx$10 times the critical plasma density, the ion density was equal to it. The simulation box had a size of $85.6~\mu$m$\times$ $112~\mu$m and consisted of $5376 \times 7008$ cells with $10$ particles of each kind per cell. The time step in the simulation was $1.84 \cdot 10^{-2}$ fs. The laser pulse at the wavelength of $800$ nm with a duration of $24$ fs was focused on the open end of the coil (see the upper left panel in Fig.~\ref{PIC_results}) to a spot with a diameter of 8~$\mu$m~($1/e^2$~width). Its peak intensity was set to $10^{21}$ W/cm$^2$. The chosen parameters correspond to those attainable by modern ultra-short pulse Petawatt systems, such as, for example, Vega-3 at the CLPU, Spain, Draco at the HZDR, Germany, or Appolon laser facility in France. 

Figure~\ref{PIC_results} presents the results, obtained for the $68$~$\mu$m wide 'copper' target with $0.8$~$\mu$m gap size. These results indicate that the action of intense laser pulse on the coil tip, as shown on the plot for $n_e$ at $t=0.00$~ps, creates a strong discharge current wave that propagates along the target perimeter (see $B_z$ at $0.36$~ps and $0.72$~ps). This is accompanied by the rise of electron density in the gap between the tip of the coil and its opposite end (see $n_e$ at $t=0.72$~ps). On the plot, corresponding to $0.36$~ps, another, weaker discharge current wave, propagating in the opposite direction towards the main wave can be also seen. It is caused by two factors: (i) some part of laser energy is absorbed by the target stalk due to reflections from the tip and the finite spatial width of the initial pulse, thus creating another discharge wave that propagates from the stalk towards the tip, and (ii) the plasma near the tip expands so quickly that it becomes conductive yet when the discharge current pulse is being formed. When the main current pulse reaches the end of the coil, it is split into two parts (see Fig. \ref{PIC_results}, $B_z$ at $1.08$~ps). The one part propagates along the stalk while the other travels trough the gap, which is already filled with the expanded plasma and is a good conductor. The passed pulse continues to travel along the coil (see Fig. \ref{PIC_results}, $B_z$ at $1.44$~ps). The simulation results for the subsequent time moments demonstrate that this process is repeated on the next cycles, although the amplitude of the current pulse decreases with each round trip.

\begin{figure}
    \centering
    \includegraphics[width = \linewidth]{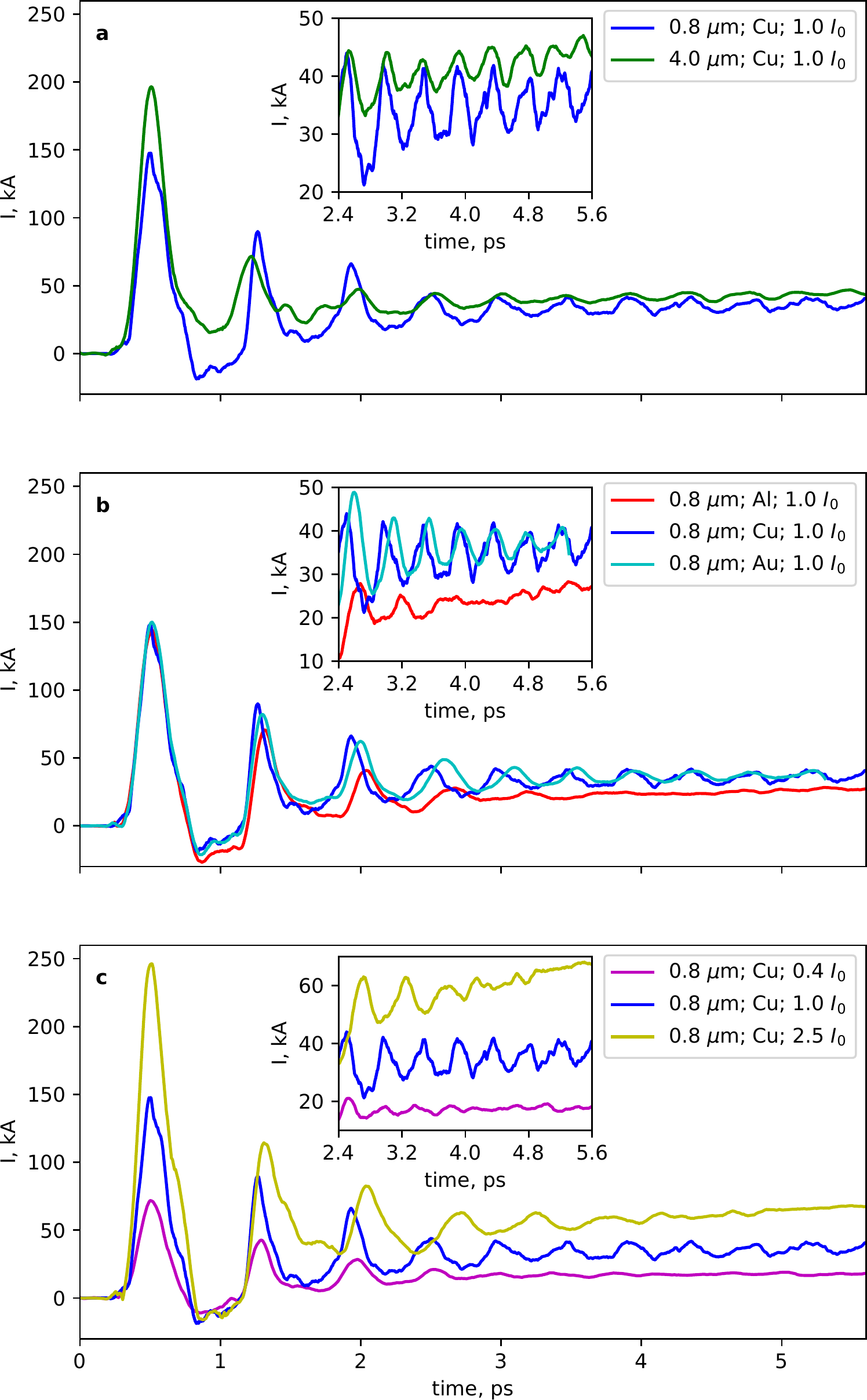}
    \caption{Total electric current in a fixed cross-section of the coil as function of time for \textbf{a} 'copper' coil targets with gap sizes of $0.8$~$\mu$m and $4.0$~$\mu$m, irradiated by laser pulses with the intensity of $I_0=1.0 \cdot 10^{21}$~$W/cm^2$; \textbf{b} coil targets, made of 'aluminium', 'copper' and 'gold', with gap size of $0.8$~$\mu$m, irradiated by laser pulses with the intensity of $I_0=1.0 \cdot 10^{21}$~$W/cm^2$; \textbf{c} 'copper' coil targets with gap size of $0.8$~$\mu$m, irradiated by laser pulses with intensities of $I_0=1.0 \cdot 10^{21}$~$W/cm^2$ and $2.5 I_0 = 2.5 \cdot 10^{21}$~$W/cm^2$. The inset panels in \textbf{a}-\textbf{c} show in more detail the time evolution of the electric current after the first 3 cycles.}
    \label{fig:comparison_I_t}
\end{figure}

In order to characterize the time evolution of the discharge current pulse for different target geometries and compositions, as well as for different intensities of the laser pulse with the PIC simulations, we considered the magnetic field, averaged over a small area at a fixed point on the coil boundary as a function of time. The value of magnetic field was then converted into total electric current in the cross-section of the coil. First, we examined two 'copper' targets with different gap sizes ($0.8$~$\mu$m and $4.0$~$\mu$m), irradiated by laser pulses with the intensity of $10^{21}$~$W/cm^2$. The obtained results, presented in Fig.~\ref{fig:comparison_I_t}, \textbf{a}, indicate that in both cases the discharge pulse is effectively trapped in the formed close circuit, so that the peak of the initial current wave is followed by the other peaks that are observed at approximately equal time intervals. The amplitude of the current pulse decreases considerably on the first 3 cycles, but then the process becomes quasi-stationary, and the pulse propagates in the loop without significant variations. The oscillations, especially at later stages after the first 3 cycles, are more pronounced for the 0.8~$\mu$m gap target, while for the $4.0$~$\mu$m gap target they quickly decay to the the amplitude about 2 times lower than that for the $0.8$~$\mu$m gap target. Such behaviour is explained by the lower electron density and lower plasma conductivity in a larger gap, and thereafter a lower part of the total current which continues annular oscillations. 

Consider the targets with 0.8~$\mu$m gap, made of 'aluminium', 'copper' and 'gold', see Fig.~\ref{fig:comparison_I_t}, \textbf{b}. Different ion masses lead to different plasma expansion, thus affecting the conditions for propagation of the discharge current pulse. As can be seen, the 'copper' and 'gold' targets exhibit similar behaviour, while the 'aluminium' target yields quite different results. The former means that the good conductance is achieved when the discharge pulse makes the first turn. In the 'aluminium' target, however, the expansion is too fast, so that a large plasma cloud is formed, leading to the damping and reflecting of the pulse. In addition, the negative effect of the opposite discharge current pulse manifests itself at the early stages of the interaction for the 'aluminium' target. 

The results for different laser pulse intensities are presented in Fig.~\ref{fig:comparison_I_t}, \textbf{c}. In this context the initially considered intensity of $I_0=10^{21}$~$W/cm^2$ appears to be reasonable for the creation of the trapped discharge current wave. Lower intensities ($4.0 \cdot 10^{20}$~$W/cm^2$) lead to insufficient expansion of the plasma, formed near the target tip, thus resulting in a considerable decrease of the amplitude of the pulse each time it passes through the gap. Higher intensities ($2.5 \cdot 10^{21}$~$W/cm^2$), on the one hand, create higher initial electric currents, but on the other hand, cause a too big dense plasma cloud preventing a lossless and dispersionless propagation of the discharge current wave. In this situation a quasi-stationary magnetic field is formed in the target self-consistently with the static current in the loop. Note, that this long-living magnetic field was created by a few fs laser pulse, demonstrating a path for using short laser pulses for a set of applications dealing with strong magnetic fields.

Consider the properties of the THz radiation, generated by 'copper' coil targets with a gap size of $0.8$~$\mu$m, irradiated by laser pulses with the intensity of $10^{21}$~$W/cm^2$. The radiated power of a current oscillating in closed round loop can be obtained analytically by calculating the vector potential from the retarded current density. In the wave zone 
\begin{equation}
    \mathbf{A}=\frac{1}{cR}\int\mathbf{I}\left(\mathbf{r},t-\frac{|\mathbf{R}-\mathbf{r}|}{c}\right)d^3r,
    \label{A_init}
\end{equation}
where $\mathbf A$ and $\mathbf I$ are is vector potential of the irradiated wave and electric current in the system respectively, $\mathbf R$ is an observation point. After Fourier decomposition with $R\gg r$ it reads 
\begin{equation}
    \mathbf{A}_\omega=\frac{e^{i k R}}{cR}\int\mathbf{I}_\omega\left(\mathbf{r}\right) e^{-i \mathbf k \mathbf r} d^3r,
    \label{A_omega}
\end{equation}
with $\mathbf k$ being the wave vector. Assuming for estimate a single short current pulse propagating along the coil $I(\omega t - \phi)$, where $\phi$ is the polar angle, for the irradiated power $\frac{d P_\omega(\Omega)}{d\Omega d\omega} =\frac{cR^2}{4\pi^2}\vert \left[\mathbf k,\mathbf A_\omega \right] \vert^2$ per solid angle $d\Omega$ we obtain
\begin{multline}
    \frac{d P_\omega(\Omega)}{d\Omega d\omega} =  \frac{a^2k^2 I_\omega^2}{8 \pi c} \times \\
    \times \left(\left(J_0(\chi)+J_2(\chi)\right)^2+\cos^2\theta (\left(J_0(\chi)-J_2(\chi)\right)^2) \right),
    \label{power}
\end{multline}
where $J_s(\chi)$ is the Bessel function of $s$-th order, $\chi = ka\sin\theta$ and $\theta$ is the angle from the coil axis direction. The emitted wave is circularly polarized, with the $J_0(\chi)$ and $J_2(\chi)$ terms relating to the opposite polarization directions.
Electric field amplitude at a distance $R$ can be estimated as:
\begin{equation}
    \label{eqn:electric_field}
    E_{\omega} = \sqrt{\frac{4 \pi c}{R^2}  \frac{dP_{\omega}}{d\Omega}}
\end{equation}

\begin{figure}
    \centering
    \includegraphics[width = \linewidth]{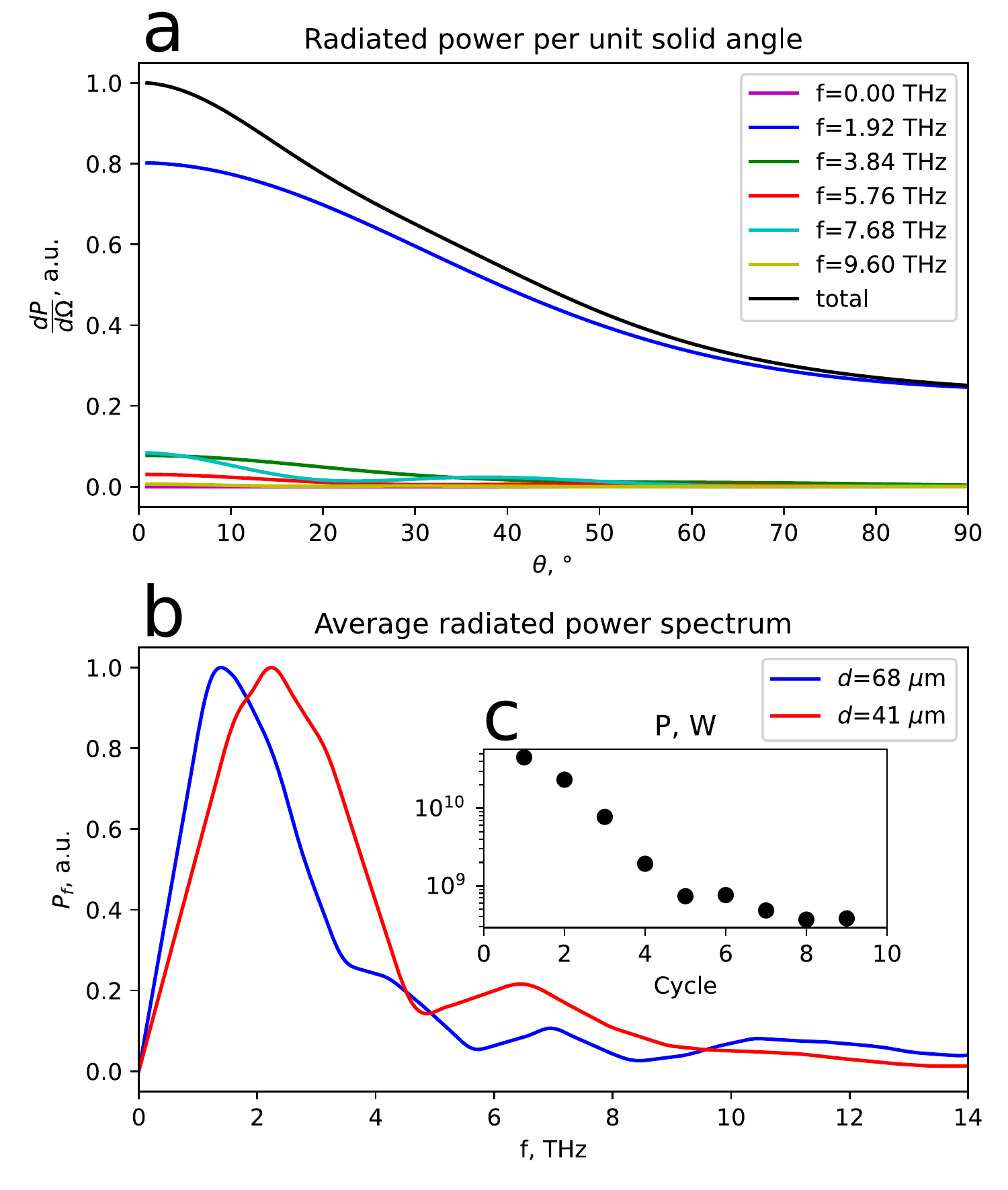}
    \caption{Analytical estimates of the properties of THz radiation, generated by coil targets, irradiated by laser pulses with the intensity of $1.0 \cdot 10^{21}$~$W/cm^2$. \textbf{a} Radiated power per unit solid angle as a function of the angle between the coil axis and the direction of measurement for different frequency components; results correspond to 'copper' target with outer diameter of 68~$\mu$m and a gap size of 0.8~$\mu$m. \textbf{b} Radiated power spectrum for 'copper' coil targets with outer diameters of 68~$\mu$m and 41~$\mu$m and a gap size of 0.8~$\mu$m; the spectrum is averaged over the first 10 and 6 cycles, respectively. \textbf{c} Average total radiated power on each cycle for 'copper' coil target with outer diameter of 68~$\mu$m and a gap size of 0.8~$\mu$m.}
    \label{fig:radiation_data}
\end{figure}

Using the estimate, the angular distribution of the radiated power per unit solid angle for the considered model setup are calculated and presented in Fig.~\ref{fig:radiation_data}, \textbf{a}. They were obtained with the current profile extracted from the PIC simulations, for a single forth oscillation when the process stabilizes, considering it as a periodic signal and using Eq.~(\ref{power}). Selecting other oscillations may lead to different results, since the shape of the pulse in simulations changes while it propagates along the coil. Moreover, the central frequency, that corresponds to the maximum of total radiated power, slowly shifts with time. This effect is mainly explained by a decrease of the time of one round trip, which changes as the inner surface plasma expands into the target cavity. In general, the angular distributions of the radiated power for other profiles are qualitatively similar to the curves in Fig.~\ref{fig:radiation_data}, \textbf{a}. The majority of the total radiated power per unit solid angle is produced at the main central frequency. The obtained angular distribution of the radiated power at the main frequency has a pronounced maximum along the coil axis. The angular distributions of the other frequency components are more complex, for some of them a larger fraction of power is radiated in the plane of the coil rather than along its axis, although the contribution to the total radiated power from them is insignificant.

To obtain the total power spectrum, the angular distributions for all frequencies over the half-space were integrated. In this case the whole signal was used, which accounts for the variations of the shape of current pulse, and the frequency shift. The obtained curve for the 'copper' coil target with the outer diameter of $d_1=68~\mu$m is shown in Fig.~\ref{fig:radiation_data}, \textbf{b}, with the blue curve. For comparison, with a separate PIC simulation a target with the outer diameter of $d_2=41~\mu$m was analyzed. The corresponding radiated power spectrum is shown with the red curve in the same panel. As can be seen, the main frequency for the second target shifts from $f_1=1.4$~THz to $f_2=2.3$~THz. The relation of the two frequencies is $\frac{f_1}{f_2} \approx 0.61$, which approximately matches the inverse ratio of their sizes $\frac{d_2}{d_1}=0.60$, since the latter define the time of one round trip. Indeed, the target size allows to control the emitted THz radiation frequency range.

The total power irradiated in the THz range in a half-space is estimated by integration the radiated power spectra over all frequencies. The obtained result is shown in Fig.~\ref{fig:radiation_data}, \textbf{c}, in logarithmic scale. Initial radiated power, produced by the current pulse over the first round, exceeds {$40$~GW}. It corresponds to the electric field amplitude of about {$0.1$~GV/cm} at the main frequency at a distance of $1$~cm from the target at its axis. An approximately exponential decay continues for several cycles, where the logarithm of the radiated power decreases linearly, and then it stabilizes. On late times of $5.6$~ps the emitted power is about {$\sim 0.4$~GW} THz radiation. The observed exponential decay is attributed to the losses near the slit and the stalk, where a part of the discharge current wave goes to the ground each oscillation round. This proposes the way to optimization for a longer multi-GW power as to modify conditions near the stalk or the stalk itself to prevent a considerable separation of the discharge wave near the mounting point. One may consider free-standing targets in this concern. 
Of course, an intense discharge pulse, propagating along an infinite wire irradiated on the free end, may also emit THz radiation \cite{Nakajima2017,Tian2017, Zhuo2017, Zeng2020}. In this situation, the problem of the circuit closure and losses at the mounting point are absent, though the radiation angular dependence and the spectrum would widen. 
Based on the presented numerical analysis, from $6.3$~J of laser energy {$\approx 0.12$~J} is re-emitted in the THz range. The conversion efficiency is then about {$2 \%$}, consistent with that reported in Ref.\cite{Zeng2020,Nakajima2017,Tian2017} for THz emission efficiency from a straight wire. In the presented setup, it is complemented with the good spectral control and directional angular distribution of the emitted radiation, which make this technique a promising tool for generation of intense THz radiation with desired properties. Further optimization of the interaction condition, geometry, material, and the mounting of the loop may result in a higher conversion efficiency. 

\begin{acknowledgments}
The work was partially supported by Ministry of Science and Higher Education of the Russian Federation (agreement \#075-15-2021-1361). We acknowledge resources of NRNU MEPhI High-Performance Computing Center. \end{acknowledgments}




\providecommand{\noopsort}[1]{}\providecommand{\singleletter}[1]{#1}%

\end{document}